# FTIR and GCMS analysis of epoxy resin decomposition products feeding the flame during UL 94 standard flammability test. Application to the understanding of the blowing-out effect in epoxy/polyhedral silsesquioxane formulations

Wenchao Zhang [a], Alberto Fina [b,*] Giuseppe Ferraro [b], Rongjie Yang [a]

a- National Engineering Technology Research Center of Flame Retardant Materials, School of Materials, Beijing Institute of Technology, PR China

b- Dipartimento di Scienza Applicata e Tecnologia, Politecnico di Torino, Sede di Alessandria, Italy

*corresponding author: alberto.fina@polito.it

**Abstract**

A novel method was developed for the sampling of volatiles produces by polymer decomposition during UL94 standard flammability tests, allowing to collect, separate and analyze the precise composition of the fuel mixture feeding the flame in the real flammability tests. The system was validated on epoxy resin/polyhedral oligomeric silsesquioxanes and found extremely informative for the understanding of the flame retardancy mechanisms of POSS, previously referred to as the "blowing-out effect". Collected products were analyzed by infrared spectroscopy and gas chromatography/mass spectroscopy, to identify the gaseous and liquid decomposition products, allowing to depict a comprehensive decomposition pathway for the epoxy resin. Lightweight volatiles, gaseous at room temperature, showed limited differences as a function of DOPO-POSS, whereas mixtures of liquid products evidenced for dramatic changes in the relative concentration as a function of DOPO-POSS. In pristine epoxy resin, the most abundant products were recognized as benzene, phenol, naphthalene and toluene, along with several tens of other aromatic products observed in lower amounts. The presence of DOPO-POSS at low concentration (2.5%) radically changes the composition of the aromatic volatiles mixture, as no significant amount of benzene is produced, while phenol becomes the main product, accounting for about half of the total, isopropyl phenol and bisphenol A, along with several tens of other products in lower concentrations. Such modification of the fuel mixture feeding the flame provided an explanation for the lower flammability of this formulation, as well as for the phenomenology of the blowing out effect.

# 1. Introduction

Epoxy resins are well-known materials that show advantageous properties such as ease of processing, low cost and good mechanical properties together with environmental advantages [1-4]. These are commonly used as advanced composite matrices in the electronic and electrical industries where a high quality flame-retardant performance is required, as the fire risk is a major drawback of these materials [5]. Traditionally, halogenated compounds have been widely used as co-monomers or additives with epoxy resins to obtain flame-retardant materials. However, flame-retardant epoxy resins containing bromine or chlorine can produce poisonous and corrosive smoke and may produce highly toxic halogenated dibenzodioxins and dibenzofurans [6-8] . Therefore, the development and application of halogen-free flame retardants has been the subject of extensive investigation in the last decade.

Phosphorus-containing compounds have shown promising application as halogen-free, flame retardants in epoxy resins [2, 9]. On one hand, the PO• radical and HPO produced by the pyrolysis of phosphorus-containing flame retardant could react with the radicals of H• and •OH, thus reducing the concentration of highly reactive radicals in the gas phase. On the other hand, P-compounds may convert into phosphoric acid during decomposition, acting in the condensed phase inducing the formation of a protective carbonaceous layer, which is highly thermally stable and can retard further decomposition of polymer chains. 9,10-dihydro-9-oxa-10-phosphaphenanthrene-10-oxide (DOPO) is a type of cyclic phosphate with a diphenyl structure, which has high thermal stability, good oxidation resistance, and good water resistance [2, 10, 11]. DOPO is a well-known phosphorus-containing FR, especially in epoxy resins, in which DOPO or its derivatives as flame retardant have been widely reported [12-15].

Polyhedral oligomeric silsesquioxane (POSS) are a kind of inorganic-organic hybrid materials which nanostructures have become attractive because of their environmental neutrality, good heat resistance as well as excellent thermoxidative stability [16-20]. Attention was directed to the use of POSS, as a great potential for producing materials characterized by improved flame retardancy along with superior physical properties [21-26]. Research results indicated that POSS exert astonishing effects on flame reaction of polymers, because they can act as stand-alone fire retardants or as synergistic fire retardants [27-30].

In a previous work [31] by some of the authors, a very peculiar behavior was observed during UL-94 test for the epoxy resin with 2.5 wt. % DOPO-POSS, which was referred to as the "blowing-out effect". Phenomenolgy of the "blowing-out effect" (Figure 1) was originally described as: "after the sample was ignited, it showed an unstable flame for several seconds; with the white gaseous products jetting outward from the condensed-phase surface, the flame was extinguished, it looks like that the gas blew out the flame" [32].

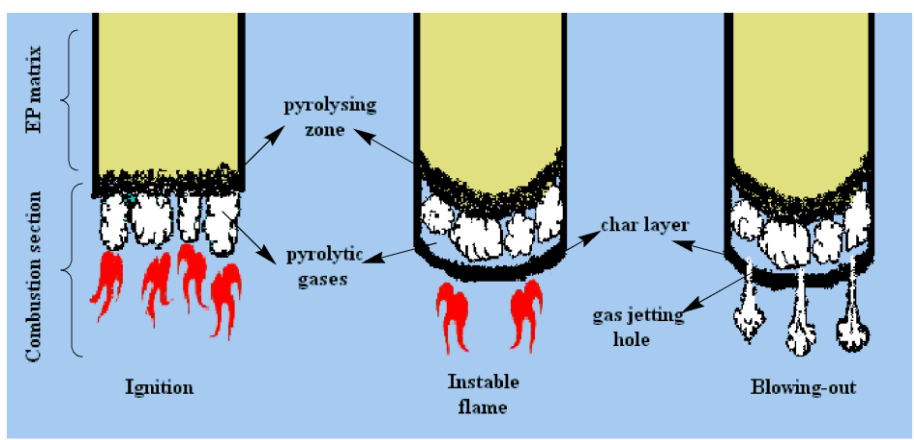

Figure 1: Schematic model of the blowing-out effect [32].

Such effect was detected in the diglycidyl ether of bisphenol A (DGEBA) cured with Diammino diphenyl sulphone (DDS which are flame retarded by DOPO-POSS [31, 33]. The mixture of POSS and DOPO was also shown to deliver the blowing-out effect, while the use POSS or DOPO alone do not have the same effect [32, 34]. The blowing-out effect originates from the confinement of gaseous mixture within closed cells formed in the condensed phase [33]. In DGEBA/DDS this can efficiently be obtained only at 2.5% DOPO-POSS, condition at which the viscosity of the decomposing condensed phase is optimized to trap decomposition gaseous products in large bubbles. Once the pressure in such gaseous chamber increases above a threshold, the cells may suddenly release a jet of volatiles, which is clearly visible during the test and previously reported [35]. Such jets resemble very much the condensed fuel released from the wick of a candle soon after the forced extinguishment of its flame and appear to contain large amounts of gaseous and/or liquid volatiles. Such volatile products are generally expected to produce a large flame in correspondence of the jets, whereas in the practical observation only small flames are typically produced towards the visible end of the jet [35]. This suggests a limited flammability of the mixture of products jetted out from the specimen, despite no experimental evidences were previously reported to support this hypothesis. This paper aims at the development of a new method to sample such jetted products and at the study of their chemical composition, in order to explain their apparent low flammability.

## 2. Experimental

### 2.1 Materials

Diglycidyl ether of bisphenol A (DGEBA, E-44, epoxy equivalent = 0.44 mol/100 g) was purchased from FeiCheng DeYuan Chemicals CO., LTD. The 4, 4'-diaminodiphenylsulphone (DDS) was purchased from TianJin GuangFu Fine Chemical Research Institute. DOPO-POSS was synthesized in our laboratory. DOPO-POSS was mixture of perfect $T_8$ cage and unperfect $T_9$ cage with one Si-OH group on it (Figure 2) [36].

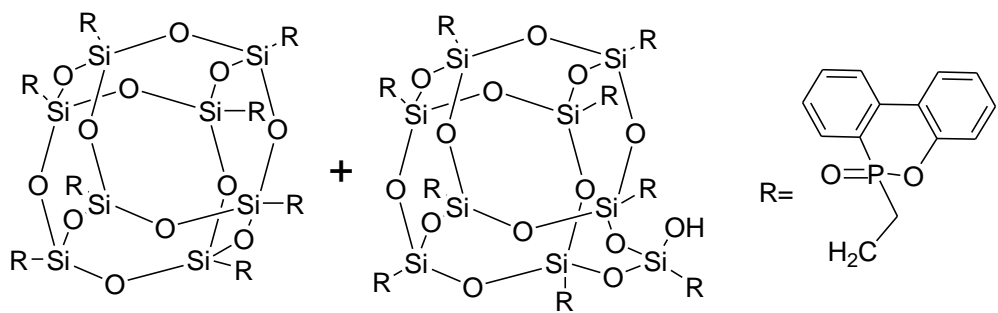

Figure 2: Typical chemical structures of DOPO-POSS molecules.

*2.2. Preparation of the cured epoxy resins*

The cured epoxy resins were obtained using a thermal curing process. At first, the DOPO-POSS was dispersed in DGEBA by mechanical stirring at 140 °C for 1 h. The mixture obtained was homogeneous and transparent. After that, the curing agent DDS was then added (weight ratio of DGEBA to DDS was 10:3). The epoxy resins were cured at 180 °C for 4h to make the curing complete. The contents of the DOPO-POSS in the EP composites were 2.5, 5 and 10 wt.%.

*2.3. Flammability test methods*

The limiting oxygen index (LOI) was obtained using the standard ASTM D 2863 procedure, which involves measuring the minimum oxygen concentration required to support top ignited candle-like combustion of plastics. An oxygen index instrument (Rheometric Scientific Ltd.) was used on samples of dimensions $100 \times 6.5 \times 3$ mm$^3$. Bottom ignited vertical burning tests were performed using the UL-94 standard on samples of dimensions $125 \times 12.5 \times 3$ mm$^3$. In this test, the burning grade of a material was classified as V-0, V-1, V-2 or unclassified, depending on its behavior (dripping and burning time), accordingly with the ASTM D 3801 standard.

*2.4. Analyses of volatiles collected during flammability test*

Volatiles obtained from sample decomposition during a UL94 flammability test were collected via a specifically designed trapping system made of a cooled U trap, in which the air flow in imposed by a downstream vacuum pump, assembled to a glass tube placed a few mm from the specimen during the UL94 test, at approx half its hight,as shown in Figure 3. A small amount of cotton fitted in the tube was found to act as an efficient filter for solid particles, to avoid these to be collected in the trap. Sampling was carried during 30-60 s, since the beginning of the flammability test, to accumulate sufficient products for analyses.

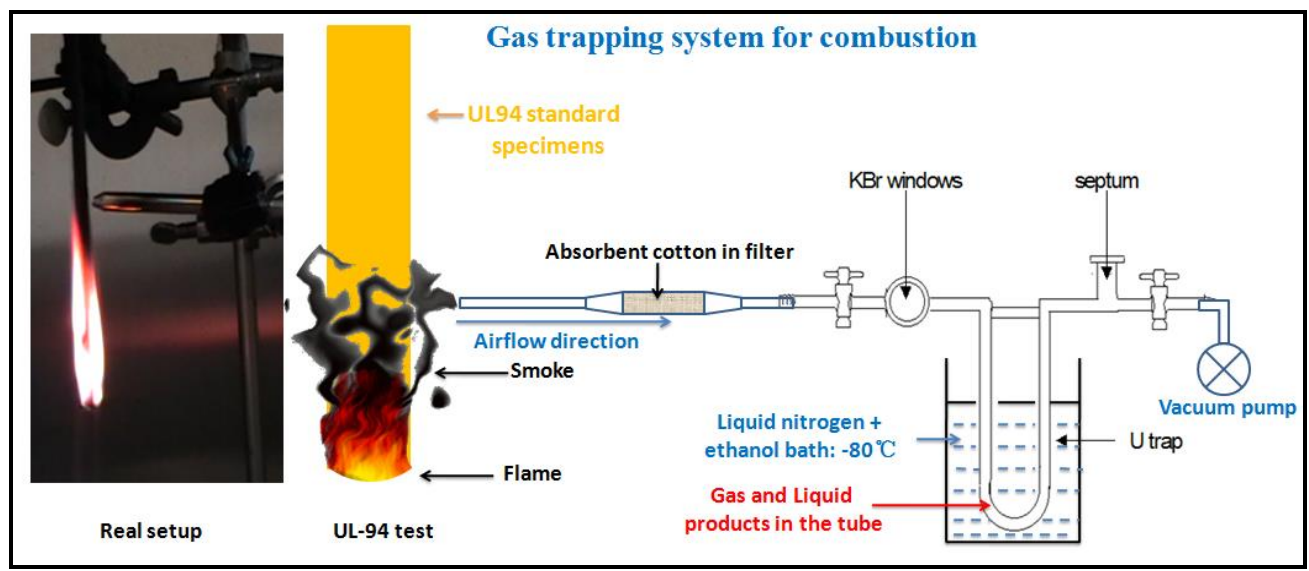

Figure 3: Gas trapping system for combustion

This system allows to collect both combustion products and polymer decomposition products that do not undergo combustion, which may clearly happen when their concentration is out of the flammablity limits. The trap was kept at -80℃, by means of a liquid nitrogen/ethanol bath, during sampling. At the end of the gas sampling time, inlet and outlet gas-tight valves on the trap were closed, the gas trapping system was placed in room temperature for about 30 min to allow gasification of the lightweight products. Then, the gases were analyzed by Fourier Transform Infrared Spectroscopy (FTIR), directly in the trap through the KBr windows on a Perkin Elmer Spectrum Two instrument. After gas analysis, the trap was opened to collect the little liquid compound, washing the U-tube with about 0.1 ml acetone (HPLC grade), to insure to collect all the products condensed on the tube inner wall. Solution collected was then analyzed by FTIR (Perkin Elmer Spectrum Two, in transmission mode), via drop-casting on a silicon chip and dried in room temperature, as well as by Gas Chromatography coupled with Mass Spectroscopy (GCMS). GCMS was performed on a Shimadzu QP2010 SE. The column used was a low polarity silica capillary column, 30 m length and 0.25 mm diameter (Varian VF1 MS, 1 µm coating). Analysis conditions were set as follows: injector temperature 200 °C, split flow 14:1, 20 ml flow controlled column fluxed with Helium (5.0 purity). Samples from epoxy and epoxy-POSS decomposition were injected as liquid (1 µl) solution in acetone (HPLC grade, ≥99.9 % purity, Sigma Aldrich). Impurities in the solvent were identified with the same method as mainly butanone and, hydroxymethyl pentanone. Solvent delay time was set at 4 minutes. The oven temperature for GC analysis was equilibrated at 30 °C, held in isothermal for 10 minutes, then raised with a heating ramp of 5°C/min to 200°C and held isothermally for 15 minutes. Mass spectra were acquired in scan mode, from 10 to 999 m/z, 70 eV, source temperature 200°C, interface temperature 220°C. All detectable peaks in the chromatograms were manually integrated. Relative abundances in the mixture were calculated based on % area. Only chemical species having relative abundances above 0.5% for at least one of the formulations tested were reported and discussed. Assignments of such peaks to the corresponding product was made based on their mass spectra, taking advantage of the library search (NIST 2014) as well as taking into consideration previous literature reports.

## 3. Results and Discussion

The UL-94 test results of the EP composites are shown in Table 1. A strong blowing-out effect (Figure S1) can be observed for EP/2.5 DOPO-POSS, whose LOI value increases from 22.0 % to 27.1 %, and the UL-94 reaches V-1 rating. With increasing of DOPO-POSS content, the blowing-out effect of EP composites weakened gradually. In the case of EP/10 DOPO-POSS, addition of 10 wt. % DOPO-POSS clearly enhanced the char formation of epoxy resin and the fire at the end of the samples developed upwards slowly after ignition, whereas the blowing-out effect disappear totally [32].

Table 1. Flame retardancy of EP/DOPO-POSS composites [33].

| Samples | Blowing-out effect | LOI (%) | UL-94 (3 mm) | $t_1$ (s) | $t_2$ (s) | Dripping |
|---|---|---|---|---|---|---|
| Pure EP | No | 22.0 | NR | burns to the clamp | / | Yes |
| EP/2.5 DOPO-POSS | Strong | 27.1 | V-1 | 8 | 5 | NO |
| EP/5 DOPO-POSS | Weak | 26.2 | NR | 44 | 35 | NO |
| EP/10 DOPO-POSS | No | 24.8 | NR | burns to the clamp | / | NO |

In order to characterize the decomposition products obtained under real combustion condition, the gas trapping system was connected to sample decomposition products released during the standard UL94 test. This aims at collecting the volatile products jetted out of the specimens showing the so-called "blowing-out effect". It is worth noting that this sampling system does acquire not only volatiles emitted in jets, but also products from volatiles oxidation in the flame (mainly water, CO, $CO_2$) as well as moisture from the ambient air. However, the organic compounds sampled with this simple system are representative of decomposition products which feed the flame, thus allowing to study the composition of the fuel mixture from the different epoxy based formulations.

Collected volatile products were first analyzed by FTIR on both gaseous and liquid products at room temperature. Then, the liquid fraction was also analyzed by GCMS to gain a more detailed description of decomposition products. In Figure 4, the FTIR spectra acquired on collected gas are reported for epoxy and epoxy/DOPO-POSS at the different concentrations. The main gaseous products evolved from EP during the combustion, which could be clearly assigned, are $CO_2$ (absorptions at 3728, 3706, 3627, 3600, 2372-2326, 2277, 720, 679-650 and 617 cm$^{-1}$); CO (absrotions at 2172, 2120 cm$^{-1}$); $SO_2$ (evidenced by the multiple peaks in the range 1340-1380 cm$^{-1}$, as well as peaks at 1165, 1138, 533 and 500 and a weak doublet around 2500cm$^{-1}$); acetylene (sharp peak at 730 cm$^{-1}$, plus its rotovibrational band observable at higher wavenumbers and doublet at 3314 and 3266 cm$^{-1}$, while its bands at about 1300 and 1360 cm$^{-1}$ are not visible, owing to overlapping with $SO_2$ signals); $CH_4$ (evidenced by the CH stretching at 3017 cm$^{-1}$ associated with the correspondent roto-vibrational modes at lower and higher frequencies as well as bending at 1305 cm$^{-1}$) and ethylene (characteristic peak at 950 cm$^{-1}$ with its rotovibrational bands, plus lower peaks barely visible at 1445 and 1890 cm$^{-1}$, while its bands around 3000 cm$^{-1}$ falls under the intense signals for methane). No signal for $NH_3$, being a common product from the decomposition of polymers containing nitrogen, were found in the gas phase. Instead, the only remaining signals after the above assignments are the broad bands at 1629 and 1600, which are perfectly compatible in position, shape and relative intensity with the spectra of nitrogen dioxide [37], which presence can indeed be explained by the oxidation of ammonia in the flame, as previously reported in the literature [38].

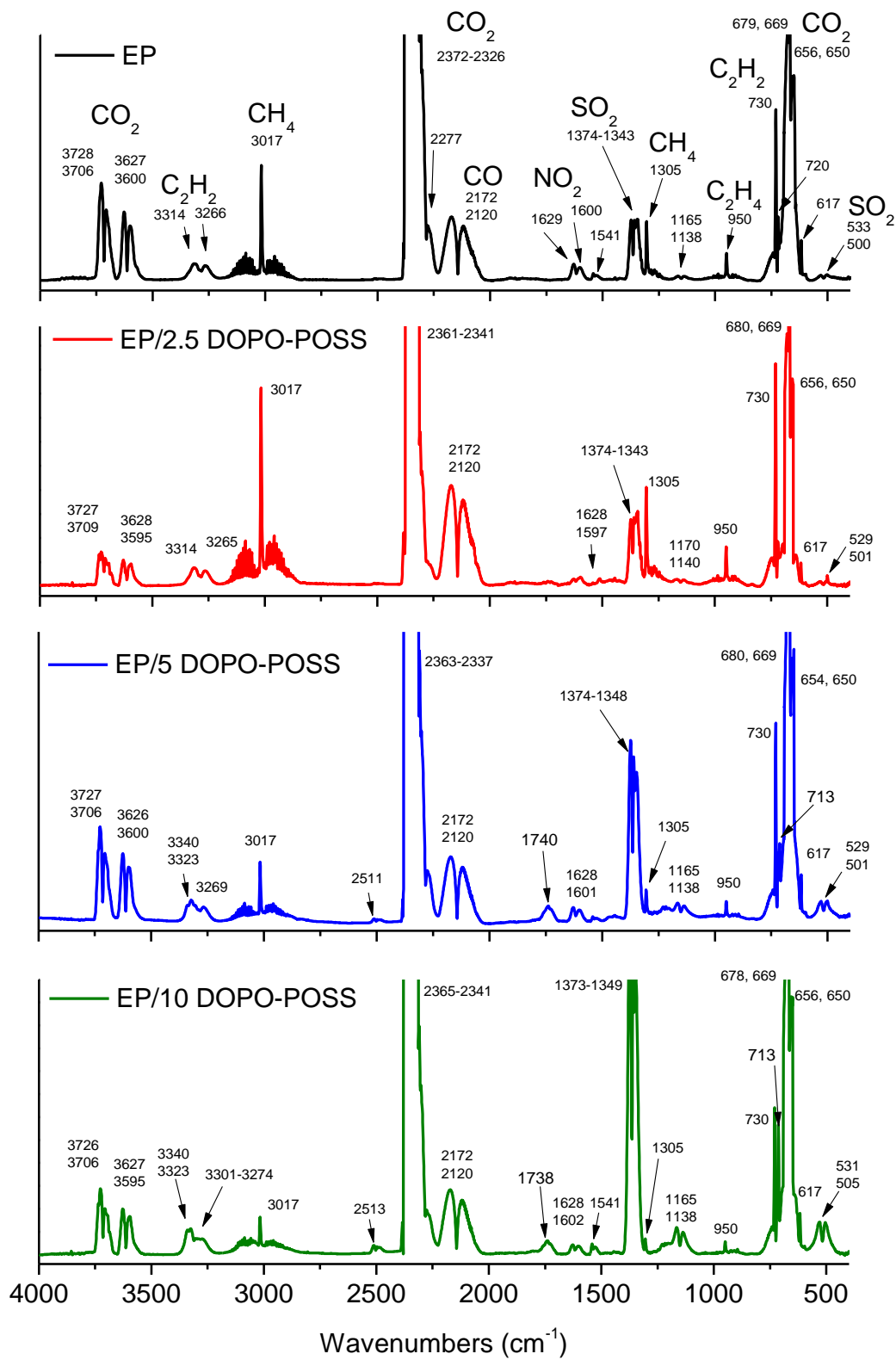

Figure 4: FTIR spectra of gas of EP and EP/DOPO-POSS sampled from the combustion unit. Magnified views of the different spectral regions are reported in Figures S2-S4.

Spectra for EP/DOPO-POSS are qualitatively very similar to that of pure EP. However, some differences in relative intensities and some additional peaks were found. In EP/2.5 DOPO-POSS, the ratios $CH_4/CO$ and $CH_4/CO_2$ are significantly higher that in pure EP, reflecting the experimental observation by which most of the fuel released from the jets did not feed the flames.

In EP/5 DOPO-POSS and EP/10 DOPO-POSS, the ratios $CH_4/CO$ and $CH_4/CO_2$ appear to be lower than for pure EP, while a significant increase of $SO_2/CO$ and $SO_2/CO_2$ is observed. Furthermore, a very sharp peak at 713 cm$^{-1}$ appears, while the shape of the acetylene doublet around 3330 cm$^{-1}$ changed, as a consequence of the overlapping of at least two new absorption bands. These two signals are likely to be assigned to the presence of HCN. Finally, a weak and broad band is observable around 1740 cm$^{-1}$, which may suggesting the presence of carbonyl groups.

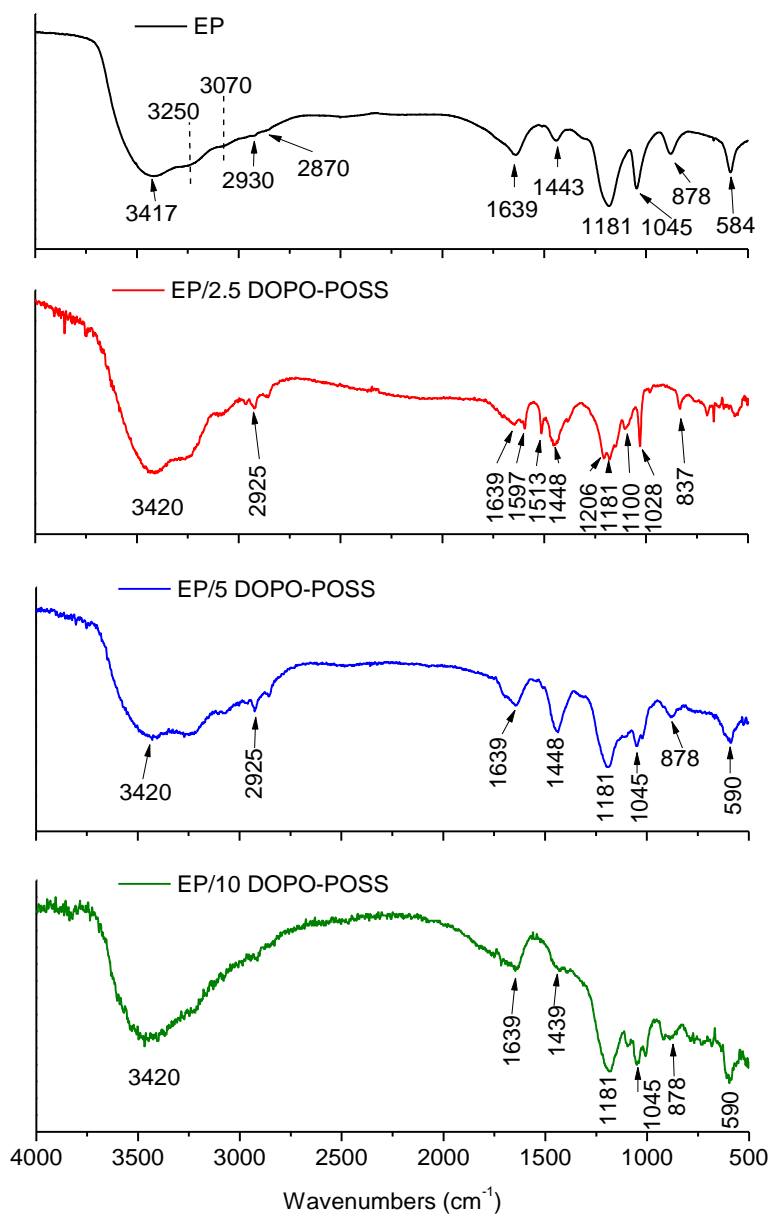

Figure 5: FTIR spectra of liquid of EP and EP/DOPO-POSS sampled from the combustion unit.

FTIR spectra of liquid decomposition products from EP/DOPO-POSS formulations are reported in Figure 5. For the pure EP, the most obvious signals are the big broad peak around 3400 cm$^{-1}$ representing OH stretching, together with the broad band at 1640 cm$^{-1}$, assigned to OH bending, thus evidencing a large amount of water. In fact, water is an obvious product in the gas phase as this is produced by the flames during the test, from epoxy dehydration in the early degradation stage and finally water is also condensed from ambient moisture. The shoulder around 3250 cm$^{-1}$ may be assigned to R-OH stretching, including phenols, while other small signals overlapped to the large OH band are visible at ~ 3070 cm$^{-1}$, suggesting aromatic C-H stretching and approx. 2930-2870 cm$^{-1}$, suggesting aliphatic C-H stretching. Beside these main signals, other broad bands are observable around 1443, 1181, 1045, 878 and 584 cm$^{-1}$, which can be assigned to several different aliphatic and aromatic C-H vibrations. For instance, band around 1440 cm$^{-1}$ is likely related to alkene CH$_2$ deformation, while regions around 1040 and 880 cm$^{-1}$ may be related to aromatic =C-H in plane/out of plane vibrations and region around 580 cm$^{-1}$ may suggest substituted aromatic ring deformations. However, precise assignment of FTIR fingerprint region is generally challenging and practically impossible in this case as the spectra correspond to mixtures of several different products, as expected from polymer decomposition. In the case of EP/DOPO-POSS, the main signals mentioned above for pure EP are confirmed. Even if some differences can be observed, especially in the spectra for EP/2.5 DOPO-POSS, showing a few additional peaks, from these FTIR spectra it is not possible to extrapolate detailed information on the differences between composition of the mixture obtained from EP and EP/DOPO-POSS. To complement the results obtained by FTIR on decomposition products from epoxy and epoxy/POSS formulations, the liquid products condensed in the U trap were analyzed by GCMS.

The decomposition products profile for pure EP is made by several tens of peaks (Figure 6), representing a number of different chemical species. In brief, the most abundant products were identified as benzene, phenol, naphthalene and toluene, on top of several other products listed in by toluene and styrene, while naphtalene is one order of magnitude lower than for pure EP. Phenol and other phenolic compounds observed at lower DOPO-POSS concentration are very low or not detectable.

Table 1. A few products could not be certainly identified, despite these appear to be assigned to dicycloaliphatic compounds, based on their mass spectra. EP/2.5 DOPO-POSS displays a very different decomposition products profile, with a maximum intensity for phenol, cumenol and bisphenol A (BPA). It is worth noting that no benzene was detected in decomposition products from EP/2.5 DOPO-POSS, while only limited amount of toluene was found. Furthermore, other phenolic compounds including cresols, ethyl phenols, Isopropenylphenol, naphtols and other minor compounds, cumulatively represent a significant fraction of the decomposition products, while the same compounds are either not detectable or found in significantly lower abundances in the decomposition products from pure EP. By comparing of chromatograms for EP and EP/2.5 DOPO-POSS it is also clear that in EP/2.5 DOPO-POSS contains products with higher elution time, which correspond to higher molar masses. These facts suggest that an overall lower fragmentation of the epoxy network was obtained in the presence of 2.5% of DOPO-POSS, thus leading to larger fragments and without the complete elimination of hydroxyl groups. Chromatogram for degradation products from EP 5/DOPO-POSS show features intermediate between pure EP and EP/2.5 DOPO-POSS. In fact, phenol remains the main product evolved, but benzene and toluene also appear in significant amounts. Higher molecular weight products, including BPA, are still detected, despite concentrations are typically lower than in the decomposition mixture from EP/2.5 DOPO-POSS. Finally, the chromatogram for decomposition product from EP/10 DOPO-POSS appears qualitatively similar to that for decomposition products from pure EP. In particular, benzene is by far the most abundant product, followed

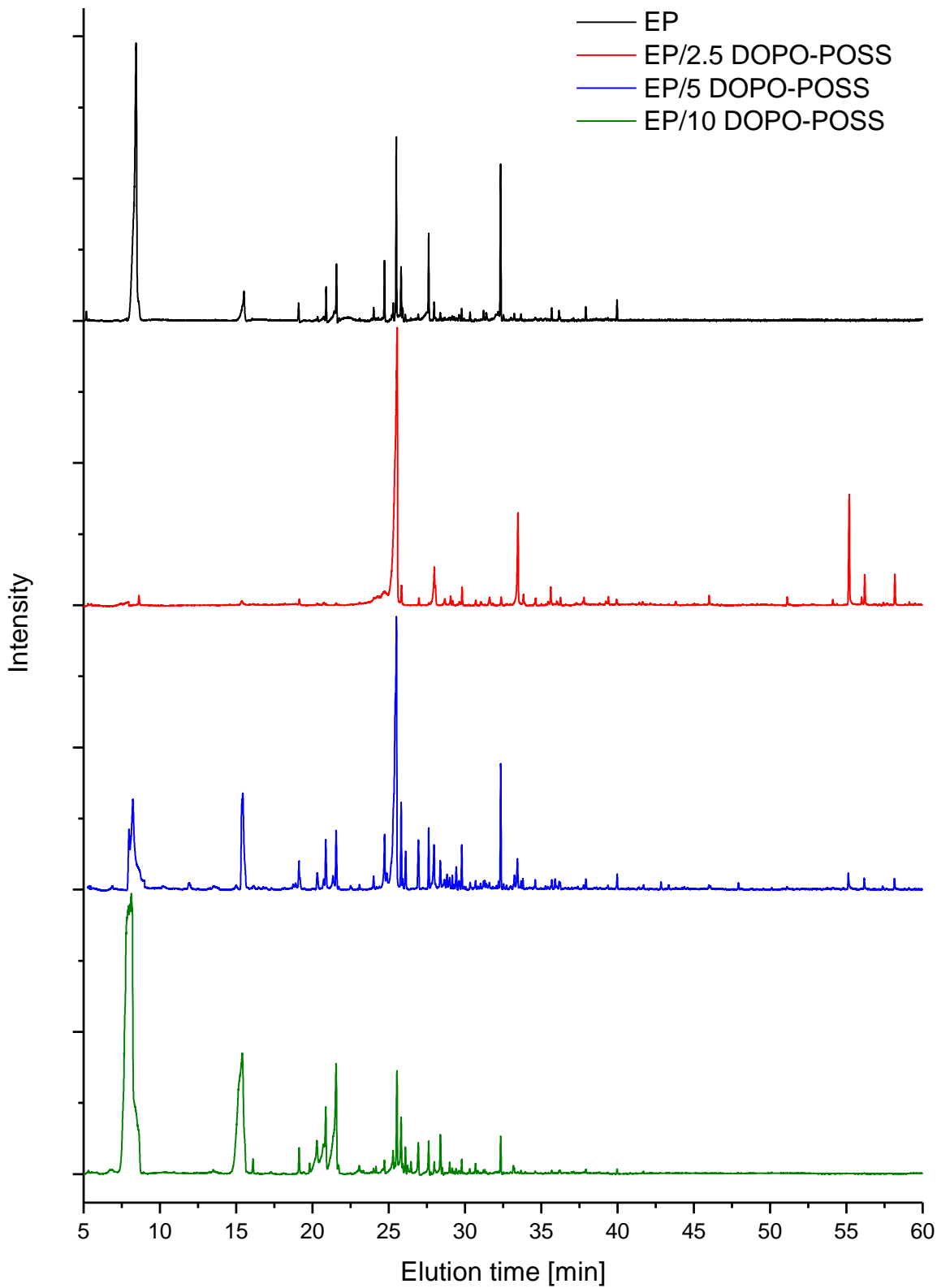

Figure 6: Chromatograms for decomposition products from pure epoxy and epoxy-POSS formulations. Normalized on the highest peak. No additional peaks detectable after 60 min elution time

Table 1: List of the main decomposition products from for pure epoxy and epoxy-POSS formulations. The table includes only the products corresponding to relative abundance ≥ 0.5% for at least one of the formulations

| # | Elution time [min] | Compound name systematic (common) | Epoxy % | 2.5% POSS % | 5% POSS % | 10% POSS % |
|---|---|---|---|---|---|---|
| 1 | 8.3 | **Benzene** | **51.1** | - | **18.4** | **60.0** |
| 2 | 15.4 | **Methylbenzene (Toluene)** | **4.8** | 0.9 | **11.1** | **17.4** |
| 3 | 20.3 | Ethylbenzene | 0.1 | 0.2 | 0.9 | 1.0 |
| 4 | 20.7 | 1,3 dimethylbenzene (m-Xylene) | 0.2 | 0.3 | 0.3 | 0.5 |
| 5 | 20.9 | Ethynylbenzene (Phenylethyne) | 1.5 | 0.1 | 1.9 | 1.7 |
| 6 | 21.6 | **Ethenylbenzene (Styrene)** | 2.9 | 0.2 | 2.3 | **6.3** |
| 7 | 24.0 | Benzaldehyde | 0.6 | 0.1 | 0.5 | 0.1 |
| 8 | 24.7 | Benzonitrile | 2.6 | 0.1 | 2.1 | 0.2 |
| 9 | 25.3 | 1-methylethenyl Benzene (α-methyl styrene) | 0.7 | * | * | 0.3 |
| 10 | 25.5 | **Phenol** | **8.3** | **54.6** | **26.7** | 2.0 |
| 11 | 25.8 | Benzofuran | 2.5 | 1.2 | 2.8 | 1.0 |
| 12 | 26.1 | 5-Norbornene-2-carboxaldehyde | 0.3 | - | 1.4 | 0.5 |
| 13 | 26.9 | Unidentified 1 (m/z=136) | 0.2 | - | 2.1 | 0.8 |
| 14 | 26.9 | Isopropyl methyl benzene (Cymene) | - | 0.5 | - | - |
| 15 | 27.6 | Indene | 3.6 | 0.1 | 2.0 | 0.8 |
| 16 | 27.9 | 2-Methyl phenol (o-cresol) | 0.7 | 3.8 | 1.8 | 0.2 |
| 17 | 28.0 | Methylamino benzene (Aniline, N-methyl) | - | 1.0 | - | - |
| 18 | 28.4 | Unidentified 2 (m/z=136) | 0.4 | - | 1.2 | 0.9 |
| 19 | 28.6 | 4-methylphenol (p-Cresol) | 0.1 | 0.5 | 0.3 | - |
| 20 | 29.0 | Dimethylamino benzene (Aniline, N,N-dimethyl) | - | 0.6 | - | - |
| 21 | 29.8 | Methyl benzofuran | 0.5 | 1.1 | 1.4 | 0.2 |
| 22 | 30.7 | 2-Ethyl phenol (o-ethylphenol) | - | 0.5 | 0.3 | - |
| 23 | 31.6 | 3-Ethyl phenol (p-ethylphenol) | - | 0.8 | 0.2 | - |
| 24 | 32.3 | **Naphthalene** | **7.0** | - | 3.9 | 0.6 |

| 25 | 32.3 | Unidentified 3 (m/z=134) | - | 0.6 | - | - |
| 26 | 33.4 | **Isopropyl phenol (Cumenol)** | - | **8.2** | 1.3 | <0.1 |
| 27 | 33.8 | Benzopyridine (Quinoline) | - | 0.5 | 0.3 | - |
| 28 | 35.6 | Isopropenyl phenol | - | 0.9 | <0.1 | - |
| 29 | 35.7 | 2-methyl naphtalene | 0.5 | - | 0.3 | <0.1 |
| 30 | 36.2 | 1-methyl naphthalene | 0.6 | - | 0.2 | <0.1 |
| 31 | 37.9 | biphenyl | 0.6 | 0.1 | 0.3 | 0.1 |
| 32 | 40.0 | biphenylene | 1.0 | - | 0.5 | 0.1 |
| 33 | 46.0 | Hydroxybiphenyl | - | 0.5 | 0.1 | - |
| 34 | 55.1 | **2,2-Bis(4-hydroxyphenyl)propane (Bisphenol A)** | - | **8.0** | 0.8 | - |
| 35 | 56.2 | Unidentified 4 (m/z=252) | - | 1.8 | 0.4 | - |
| 36 | 58.1 | Unidentified 5 (m/z=266) | - | 1.8 | 0.3 | - |

- = undetectable
* = α-methyl styrene detectable but difficult to quantify, owing to overlapping with phenol

The degradation products profile briefly discussed above clearly evidence there is no linear trend for the changes in chemicals abundances with increasing DOPO-POSS concentration. Instead, the presence of DOPO-POSS in low concentration (2.5%) is optimal to reduce the concentration of low molecular weight aromatic hydrocarbons produced during decomposition upon the application of a small flame to the specimen. When increasing the concentration of DOPO-POSS, the concentration of aromatic hydrocarbons increases again in the decomposition mixture, while reducing the phenolic compounds. The maximum effect obtained in modifying gaseous mixture at 2.5% DOPO-POSS in indeed in excellent agreement with the peculiar behavior observed during UL94 test with epoxy containing DOPO-POSS, in which a minimum flammability was observed at 2.5% DOPO-POSS concentration. Thus, the present results demonstrate that the low flammability of the volatiles jets emitted from the specimens and previously referred to as the "blowing-out effect" [31] is related to a significant change in volatile composition obtained at this DOPO-POSS concentration. In particular, the substitution of benzene and toluene with phenol and other phenolic compounds is recognized as the main responsible for the low flammability of the gaseous mixture jetted out of the specimens during UL94 test. In fact, phenol has values for flash point and auto ignition temperature (79 and 715°C, respectively) which are significantly higher than for benzene (-11 and 580°C, respectively). This implies that a mixture rich in phenol jetted out of the specimen is poorly flammable and therefore not prone to easily produce a large flame. Furthermore, difference in the boiling point (80 and 182 °C for benzene and phenol, respectively) may also play a role: once phenol is jetted out it is expected to rapidly cool in air and condense in small droplets, which may contribute to make jets clearly visible by the naked eye. In order to produce a flammable mixture, phenol in such droplets should re-evaporate, which require the associated enthalpy of vaporization (59 KJ/mol for phenol), and diffuse with oxygen, with an overall decrease of the oxidation rate. Beside phenol, being the most important compound, other phenolic compound released in the decomposition may contribute in further lowering the mixture flammability, accordingly to their relatively high flash points and auto-ignition temperature. A more comprehensive collection of flammability parameters for the main products evolved is reported in Table S1. In the flaming combustion test of EP/5 DOPO-POSS, a weaker blowing-out effect and a worse overall performance in UL94 test were observed and reported [31]. In terms of released volatiles, the lower concentration of phenol as compared to aromatic hydrocarbons may at least partially explain this behavior. When increasing the DOPO-POSS concentration to 10%, aromatic hydrocarbons are again representing the vast majority of the volatile mixture and the performance in flammability test further worsens compared to EP/5 DOPO-POSS.

Beside the evident effect of the volatile mixture composition on the flammability performance, the reasons behind the modification of the mixture composition is to be carefully examined. It is known and was previously reported [39] that for EP/2.5 DOPO-POSS the temperature in the condensed phase during a flammability test is significantly lower than in pure EP and EP/10 DOPO-POSS. In such lower temperature condition, a lower extent of epoxy resin chain scission is expected, leading to volatile chain fragments with higher molecular weight. It is also well know from previous literature reports that the decomposition occurring at lower temperature leads to high concentration of phenolic compounds, as these products are obtained from primary scission of the aromatic epoxy chains and are not further decomposed owing to the limited temperature and time available for conversion from phenols to hydrocarbons. Therefore, there is certainly an effect of temperature in determining the volatile mixture composition, demonstrated by the direct correlation between concentration of aromatic hydrocarbons and temperature of the condensed phase during the flaming combustion test. In addition to the effect of temperature, there may also be a chemical effect related to the presence of DOPO in the decomposing epoxy. GCMS analyses evidenced the presence of ethyl DOPO in the volatiles collected from EP/2.5 DOPO-POSS and EP/5 DOPO-POSS, by a small peak in the chromatogram at 57.4 min, while the same compound was not detectable in EP/10 DOPO-POSS (Figure S6). As no other phosphorous-containing species were identified in the sampled volatiles, we can conclude that, during the UL94 flammability test, at least a fraction of the DOPO in the condensed phase is released form POSS by the breaking of Si-O-C bond with no further molecule scission. The presence of DOPO in the volatiles mixture might have a role in stabilizing the phenols within the gas bubbles in the condensed phase by a radical scavenging effect and certainly has a role in the gas phase , contributing in slowing down the combustion of the phenol-rich mixture, thus eventually leading to the rapid self-extinguishment of flames observed in the UL94 test.

Based on the experimental results presented above, the decomposition mechanisms for the epoxy network can be described. The initial decomposition step of DGEBA type epoxy/amine network are quite well known form decades,[40-45] by the dehydration by the elimination of aliphatic hydroxyls, leading to C-C double bonds formation in two different positions, as depicted in Figure 7. Dehydrated structure further undergoes degradation by the scission of bonds along the chain.

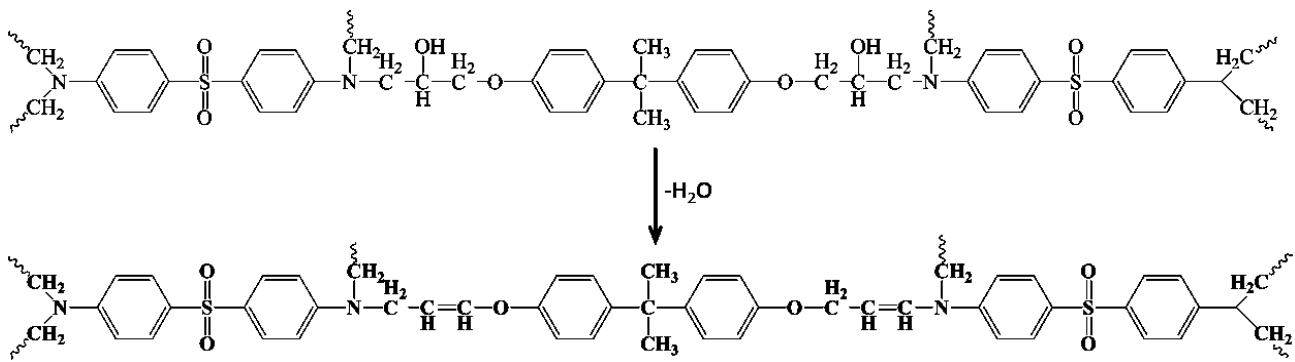

Figure 7: Dehydration of epoxy resin. Double bonds may be formed in α or β position as shown on the left and right side, respectively

Despite in controlled and slow heating the most likely scission is assumed to occur in bisphenol A at the carbon/phenyl bond, several different scissions remain possible, especially when the heating rate is high as a consequence of flaming ignition of the polymer. In DDS cross-linked epoxy, radical chain scission was shown to occur at sulphur-phenyl bond, with subsequent hydrogen abstraction by the phenyl radical and reorganization of phenylsulphonyl radical by the release of $SO_2$ [11] or formation of a polysulfone [46]. Subsequent reactions following primary scissions of the dehydrated epoxy, including further scission and reorganizations are not well known and strictly depend on the thermal scenario at which the decomposition occurs. To the best of the authors' knowledge, no previous reports are available on the detailed DGEBA-DDS resin decomposition pathways leading to volatiles feeding the flame during a flammability test. Therefore, a comprehensive yet simplified epoxy decomposition pathway was drafted and reported in Figure 8, based on the several products evidenced by FTIR and GCMS analyses discussed above.

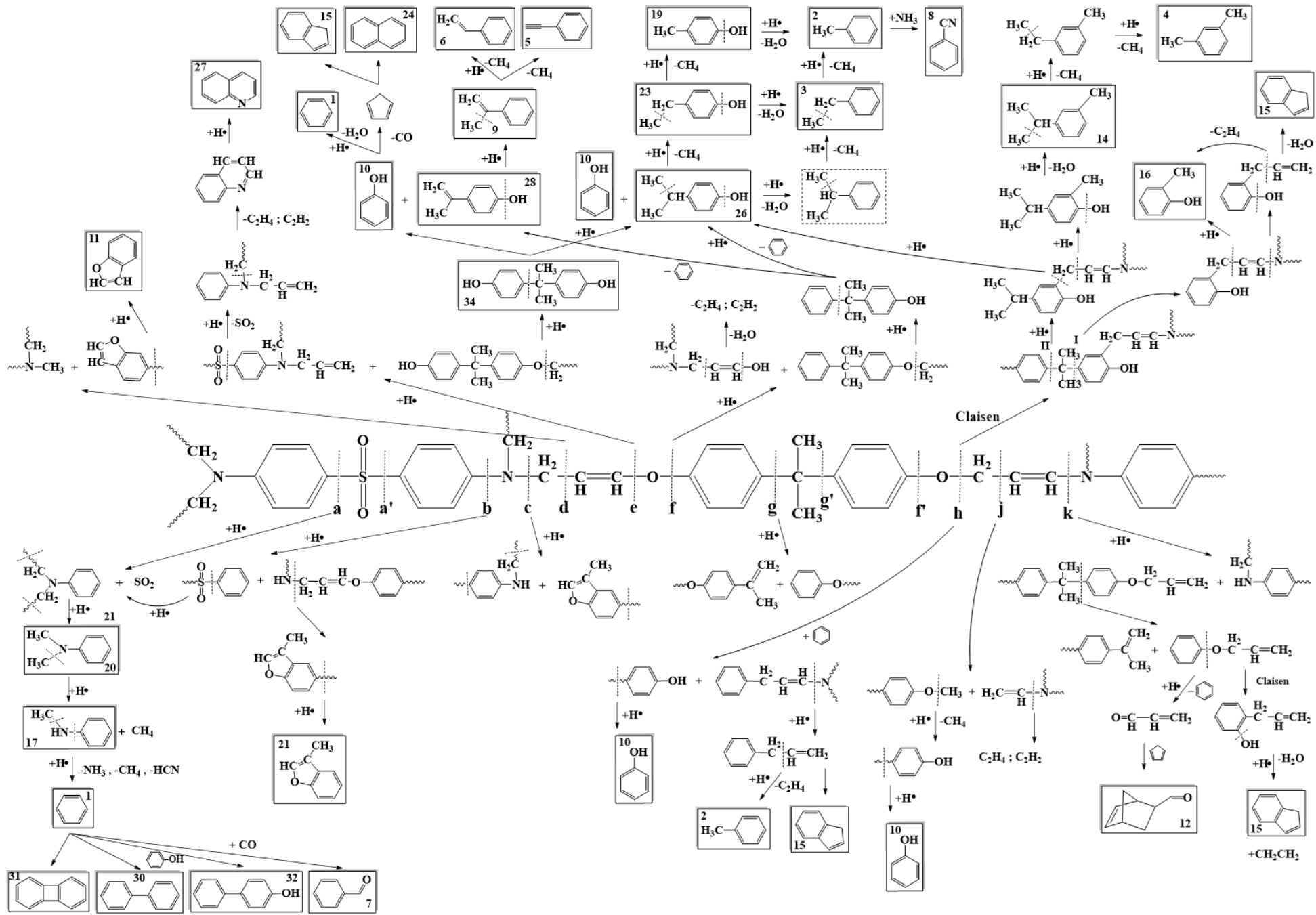

Figure 8: Decomposition pathways for DGEBA-DDS resin

Possible chain scissions are marked in letters (**a**, **b**, ..., **k**) from left to right on the structure of dehydrated DGEBA-DDS resin, regardless their differences in activation energies. In reality, reactions with lower activation energies are clearly expected to occur at a higher rate, while higher activation energies scissions require more time to lead to measurable concentrations of decomposition products. Therefore, among the possible primary chain scissions, some will clearly occur at higher rate, producing oligomers which will be later decomposed to volatiles. However, accounting for decomposition pathways from the primary chain allows to provide a simple and representative explanation of the decomposition mechanism leading to the volatile products experimentally observed. Indeed, every scission (**a**, **b**, ..., **k**) produces fragments which may undergo subsequent reactions, as briefly commented in the following.

Sulphur-phenyl bond scission (**a** and **a'** in Figure 8) leads to a phenyl radical and phenyl sulphonyl radical, which further decompose at least partially to evolve $SO_2$, as evidenced by FTIR in the gaseous products. It is also possible that part of the phenyl sulphonyl radicals reacts onto another aromatic ring to produce a polysulfone [46] producing no volatiles and remaining in the condensed phase, where it may contribute to the charring of the epoxy resin. Amino-substituted phenyl species are likely to be produced in the presence of H· radicals and alkyl radicals on the amine are likely to progressively decompose to produce methyl aniline and dimethyl aniline, which were indeed observed by GCMS in the decomposition products of EP/2.5 DOPO-POSS. The fact that substituted anilines are not detected in both pure EP and EP/DOPO-POSS at higher concentrations suggest these products to be unstable at high temperature, evolving methane and/or hydrogen cyanide clearly observed by FTIR in the evolved gases, and benzene, the latter being by far the most abundant product in the decomposition of pure EP. The production of ammonia is also postulated, despite no clear evidence of this gas was found in the FTIR, suggesting this product to be unstable and further evolving via oxidation to nitrogen and nitrogen oxide [38], particularly $NO_2$ as from FTIR assignment, or reaction with other decomposition products as detailed in the following. As the scission of the aromatic amine is expected to be radicalic, highly reactive species such as phenyl radicals and benzyne are likely produced, accounting for side reactions, *e.g.* by radical recombinations. For instance, biphenylene was observed from pure EP, EP/5 DOPO-POSS and EP/10 DOPO-POSS, which is explained by the reaction at relatively high temperature between benzene and benzyne [47]. Similarly, the formation of biphenyl, observed from decomposition of all the different formulation studied in this paper, is explained by the recombination of two phenyl radicals or, again, by the reaction of benzene and benzyne [47]. Furthermore, reaction of benzene radicals with phenol and CO (which production is explained in the following) may produce hydroxy-biphenyl and benzaldehyde [48], respectively.

Primary scission of the dehydrated epoxy in position **b** lead to a sulfoxide substituted phenyl radical, likely to evolve by further scission at the S-phenyl bond to produce $SO_2$, benzene and a substituted aniline undergoing decomposition as described above in route **a**. On the other side, the aliphatic secondary amine is prone to further decomposition at the C-N, producing a primary amine and a methyl benzofuran terminated chain, further decomposing to methyl benzofuran, always detected in the decomposition mixture in limited amounts.

Scission in position **c**, leads to an alkyl-aryl secondary amine, further decomposing to produce volatiles previously mentioned in route **a**, and a methylbenzofuran terminate chain, equivalent to route **b**.

Scission in position **d** leads to a tertiary amine, further decomposing similarly to other aliphatic amines previously mentioned, and a benzofuran terminated chain, eventually releasing benzofuran upon further scission in the presence of hydrogen radicals.

Scission in position **e** produce a propenyl substituted tertiary amine and a phenol terminated chain. The tertiary amine may decompose releasing ethylene and/or acetylene, accordingly with FTIR evidences, or cyclize to produce benzopyridine, observed in small amounts in decomposition products from EP/2.5 DOPO-POSS and EP/5 DOPO-POSS. It is worth mentioning that highly reactive ethylene and acetylene are expected to have an important role in the production of several polyaromatic hydrocarbons. Indeed, ethylene combustion is well known to produce a number of aromatics and polyaromatics, including benzene, toluene, indene, styrene, ethyl benzene, phenyl acetylene, naphthalene, methyl naphatlene [49, 50], which are all found in the sampled mixtures. On the other hand, the phenol terminated counterpart decomposition at the carbon-oxygen bond produce BPA is supported by the experimental evidences. Indeed, BPA was observed in large amount from decomposition of EP/2.5 DOPO-POSS, whereas a lower concentration was observed from EP/5 DOPO-POSS and no evidence of that was found in decomposition mixture from pure EP and EP/10 DOPO-POSS, suggesting BPA to be unstable at high temperature and further decompose at the C-phenyl bond, to produce i) phenol and para isopropenylphenol or ii) phenol and para isopropyl phenol, depending on the availability of H· radicals. Phenol is the major decomposition product from EP/2.5 DOPO-POSS and still one of the most abundant product for pure EP and EP/5 DOPO-POSS. The concentration of phenol reflects the temperature at which decomposition occur as a consequence of flaming ignition. In particular, the concentration is maximized in the case of EP/2.5 DOPO-POSS, in which decomposition temperature was demonstrated to be lower. This suggest phenol is not favored at high temperature and other compounds are obtained, either by further decomposition of phenols at higher temperature or by alternative decomposition routes. In addition to the dehydroxylation of phenol to benzene, a second route is also proposed here, by the scission to phenoxy radicals and further decomposition to produce CO and cyclopentadiene, a previously reported in the literature [51]. Cyclopentadiene is a highly reactive species at high temperature and was not detected by neither FTIR or GCMS measurements; however, the formation of some compounds observed experimentally by GCMS is straightforwardly explained by well-known reactions of cyclopentadiene. For instance, the formation of naphthalene, one of the main decomposition products in pure EP is explained by the dimerization of cyclopentadiene, while indene was also reported to be obtained from cyclopentadiene and cyclopentadienyl radicals [52]. Accordingly with the generally agreed mechanism reported in the literature [42], the expected product from BPA scission is p-isopropenyl-phenol, which is indeed found in EP/2.5 DOPO-POSS. Once again, this product appears to be unstable at higher temperature, as it is not observed in significant concentrations in decomposition products form other formulation, while by-products from its dehydroxilation, including α-methyl styrene, styrene and phenylethyne, were always found. Furthermore, p-isopropylphenol is also obtained in large amount from EP/2.5 DOPO-POSS, which formation is possible in the presence of hydrogen radicals during BPA scission. Further decomposition of isopropyl phenol via methane abstraction account for the formation of p-ethyl phenol and p-methyl phenol, while their dihydroxylation explains the formation of isopropyl benzene (observed in traces), ethyl benzene and toluene. The latter product is of particular importance in terms of concentration in the mixtures, especially in EP/DOPO-POSS at high loading, as well as a precursor for the production of benzonitrile, as a product of the well-known reaction of toluene with ammonia [53]. However, it is worth noting benzonitrile may also be obtained by the high temperature reaction of Benzene o biphenyl with hydrogen cyanide [54, 55].

Scission in position **f** produce a vinyl alcohol terminated chain, prone to further scission to produce insaturated $C_2$ hydrocarbons (ethylene and acetylene) and a derivative of BPA, undergoing a decomposition equivalent to that described in route **e**. Similar decomposition pathway is also expected by scission in position **f'**.

Scission in position **g** and **g'** may produce isopropenyl terminated and phenyl terminated chains, which may undergo further scission to evolve styrene and benzene. Scission in position **h** may also undergo chain transfer, either onto another aromatic ring or on the same aromatic ring, via Claisen rearrangement [56]. For instance, transfer to benzene and subsequent scission at the C-N bond and cyclization provides another possible route for the formation of indene, while elimination of acetylene to produce toluene remains possible. In case of Claisen rearrangement on the primary chain, an orto-substituted phenol is obtained, followed by the two possible C-phenyl scissions previoulsy discussed for BPA. On the one hand, scission in position I a monosubstituted phenol is obtained, followed by further decomposition directly to o-methyl phenol, always detected in the decomposition mixtures, or to allyl-phenol as an intermediate for methyl phenol or for indene after water elimination. On the other hand, scission in II leads to a di-substituted phenol, undergoing scission to isopropyl phenol or to methyl isopropyl phenol. However, negligible concentration of di-subtituted phenols in the decomposition mixture suggests dehydration to methyl isopropyl benzene and eventually decomposition m-xylene, always observed in small amounts, and m-methyl ethyl benzene, detected in low amounts in EP/2.5DOPO-POSS.

Scission at position **j** produces a methoxy phenyl terminated chain, which is assumed to simply convert to a phenol upon methane release, while the insaturated alkyl counterpart may evolve to ethylene and/or acetylene. Finally, scission at position **k** produces an alkyl-aryl amine (equivalent to route **c**) and a propenoxy terminated chain, which may either decompose at the phenyl-oxygen bond to release propenal or to undergo Claisen rearrangement. Despite propenal was not detected as such, experimental evidences from GC-MS analysis suggest its reaction with cyclopentadiene to produce 5-norbornene-2-carboxaldehyde, which is a clearly observed decomposition product for pure EP and EP/DOPO-POSS except at 2.5% POSS. On the other hand, allyl phenol may also be obtained by Claisen rearrangement, followed by dehydration and cyclization to produce indene.

The decomposition pathways briefly discussed above explain the formation of the main decomposition products experimentally observed. However, hundreds of competing reactions are possible in the mixture and the proposed decomposition scheme is not intended to be exhaustive of all the possible routes for the formation of volatiles. In addition, it is worth mentioning that all of the substituted aromatics hydrocarbons and phenols discussed above may in principle decompose further to produce benzene, even if this was not described all the times for the sake of brevity.

**Conclusions**

A novel method was developed and validated for the sampling of volatiles produces by polymer decomposition during UL94 standard flammability tests, allowing to collect, separate and analyze the precise composition of the fuel mixture feeding the flame in the real test. This represent a significant advance compared the traditional study of the products obtained during thermal decomposition in controlled conditions, which may not be representative of the decomposition mechanism occurring during exposure to a flame. The method was applied to both pristine epoxy resin (DGEBA-DDS) and epoxy resin containing DOPO-POSS in variable concentration, with the aim of elucidating the effect of DOPO-POSS, previously reported to dramatically affect the flammability of epoxy resin, via the so-called "blowing-out effect". Collected products were analyzed by infrared spectroscopy and gas chromatography/mass spectroscopy, to identify the gaseous and liquid decomposition products. Lightweight volatiles, gaseous at room temperature, showed limited differences as a function of DOPO POSS, the main products being $CO_2$, CO, $CH_4$, $C_2H_2$, $C_2H_4$ and $SO_2$. On the other hand, mixtures of liquid products evidenced for dramatic changes in the relative concentration as a function of DOPO POSS. In pristine epoxy resin, the most abundant products were recognized as benzene, accounting for about half of the total, phenol, naphthalene and toluene, along with several tens of other aromatic products observed in lower amounts, which contributed to depict a comprehensive decomposition mechanism for the DGEBA-DDS resin. The presence of DOPO-POSS at low concentration (2.5%) radically changes the composition of the aromatic volatiles mixture, as no significant amount of benzene is produced, while phenol becomes the main product, accounting for about half of the total, isopropyl phenol and bisphenol A, along with several tens of other products in lower concentrations. Such modification of the fuel mixture feeding the flame provided an explanation for the lower flammability of this formulation, as well as for the phenomenology of the blowing out effect. Furthermore, the compositions of the volatile mixtures from epoxy resins containing higher DOPO-POSS concentrations were found to progressively resemble the composition of pristine epoxy with increasing DOPO-POSS concentration. These facts suggested the blowing out effect, which is maximum at 2.5% POSS, to be related to a physical thermal shielding effect rather than to a chemical action of the DOPO-POSS.

Based on the detailed study of both gaseous and liquid products experimentally observed, a comprehensive yet simplified DGEBA-DDS decomposition pathway was proposed, which may pose the basis for the design of novel flame retardant additives for epoxy resins.

## Acknowledgements

This project was funded in part by International Science and Technology Cooperation Program of China (No. S2014ZR0465), National Program on Key Research Project (No. 2016YFB0302101), and National Natural Science Foundation of China (No. 51603011).

Prof Giovanni Camino at Politecnico di Torino is gratefully acknowledged for the continuous discussion and suggestions on the interpretation of decomposition pathways.

## Authors' contribution

W. Zhang carried out materials preparation, testing and contributed to data treatment and analysis. A. Fina designed the experiments and the sampling system, interpreted FTIR and GCMS results and drafted the decomposition pathways. G. Ferraro carried out GCMS analyses and participated to the discussion of decomposition reactions. R. Yang contributed to the discussion and interpretation of the results. The manuscript was mainly written by A. Fina and W. Zhang.

# SUPPORTING INFORMATION

# FTIR and GCMS analysis of epoxy resin decomposition products feeding the flame during UL 94 standard flammability test. Application to the understanding of the blowing-out effect in epoxy/polyhedral silsesquioxane formulations

Wenchao Zhang [a], Alberto Fina [b,*] Giuseppe Ferraro [b], Rongjie Yang [a]

a- National Engineering Technology Research Center of Flame Retardant Materials, School of Materials, Beijing Institute of Technology, PR China

b- Dipartimento di Scienza Applicata e Tecnologia, Politecnico di Torino, Sede di Alessandria, Italy

*alberto.fina@polito.it

1. **Flammability of epoxy/DOPO-POSS**

Blowing-out effect is clearly observable on EP/2.5 DOPO-POSS. Figure S1 shows snapshots from a flammability test video in the last seconds prior to self-extinguishment of EP/2.5 DOPO-POSS. The poorly flammable pyrolytic gases are jetted out very drastically on the char layer, and the flame is consequently pushed far from the specimen surface (Blue and red arrows). When the gases jetting is strong enough, the flame was extinguished immediately.

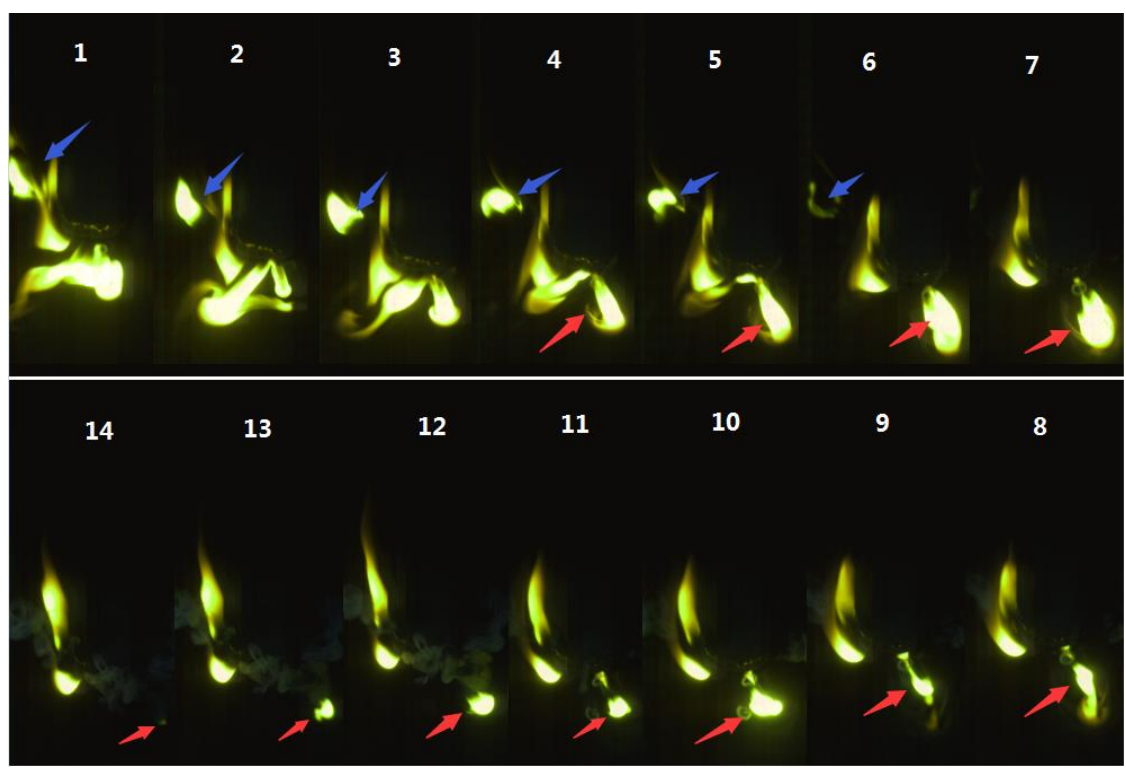

Figure S1: Extinguishing process during the last seconds of high-speed video for EP/2.5 DOPO-POSS.

**2- Analyses of evolved products**

Magnification of gas phase FTIR spectra are reported hereunder for ranges 3900-2700 cm$^{-1}$ (Figure S2), 2700-1450 cm$^{-1}$ (Figure S3) and 1450-400 (Figure S4).

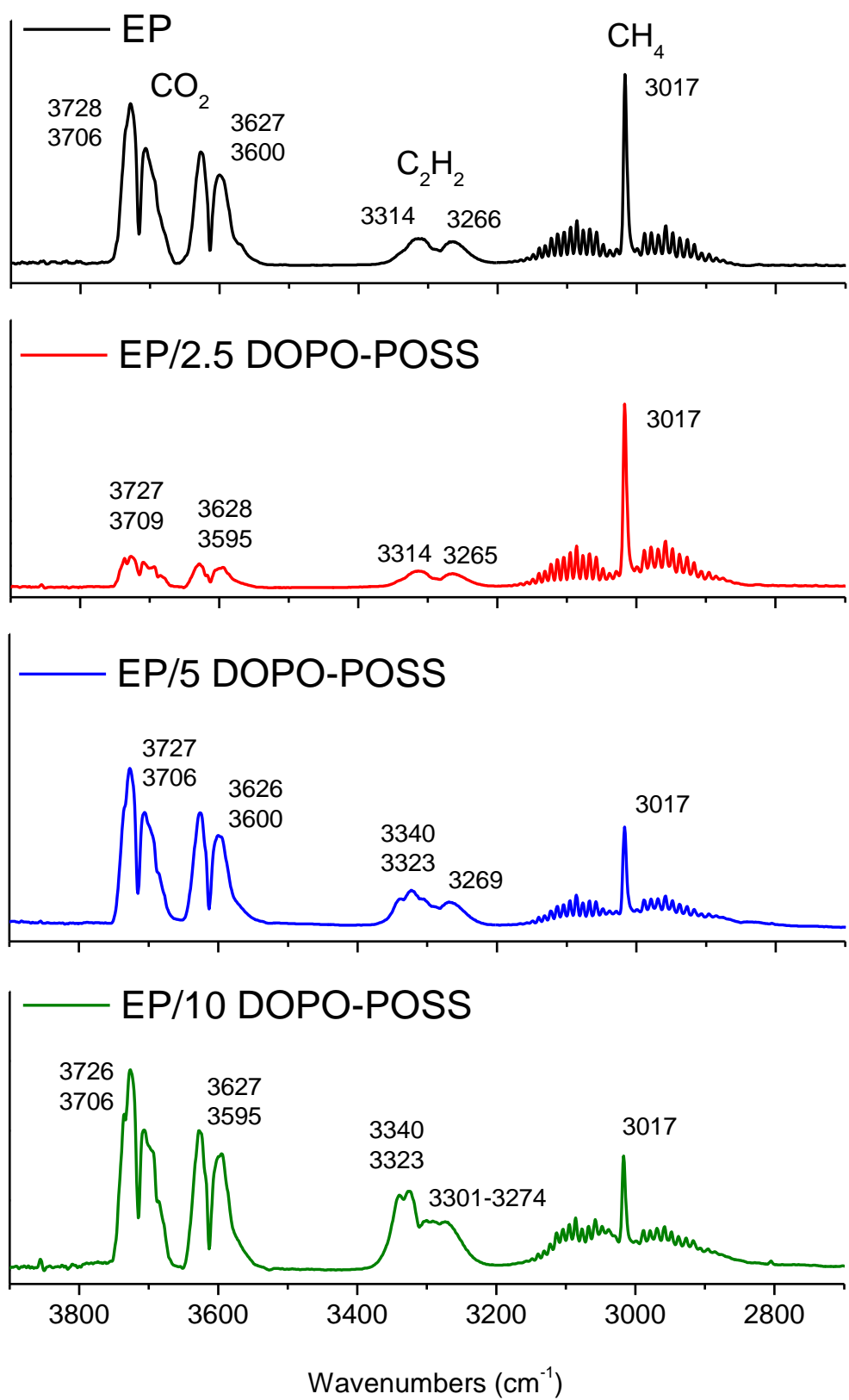

Figure S2: FTIR spectra of gas of EP and EP/DOPO-POSS sampled from the combustion unit: region 3900-2700 cm$^{-1}$

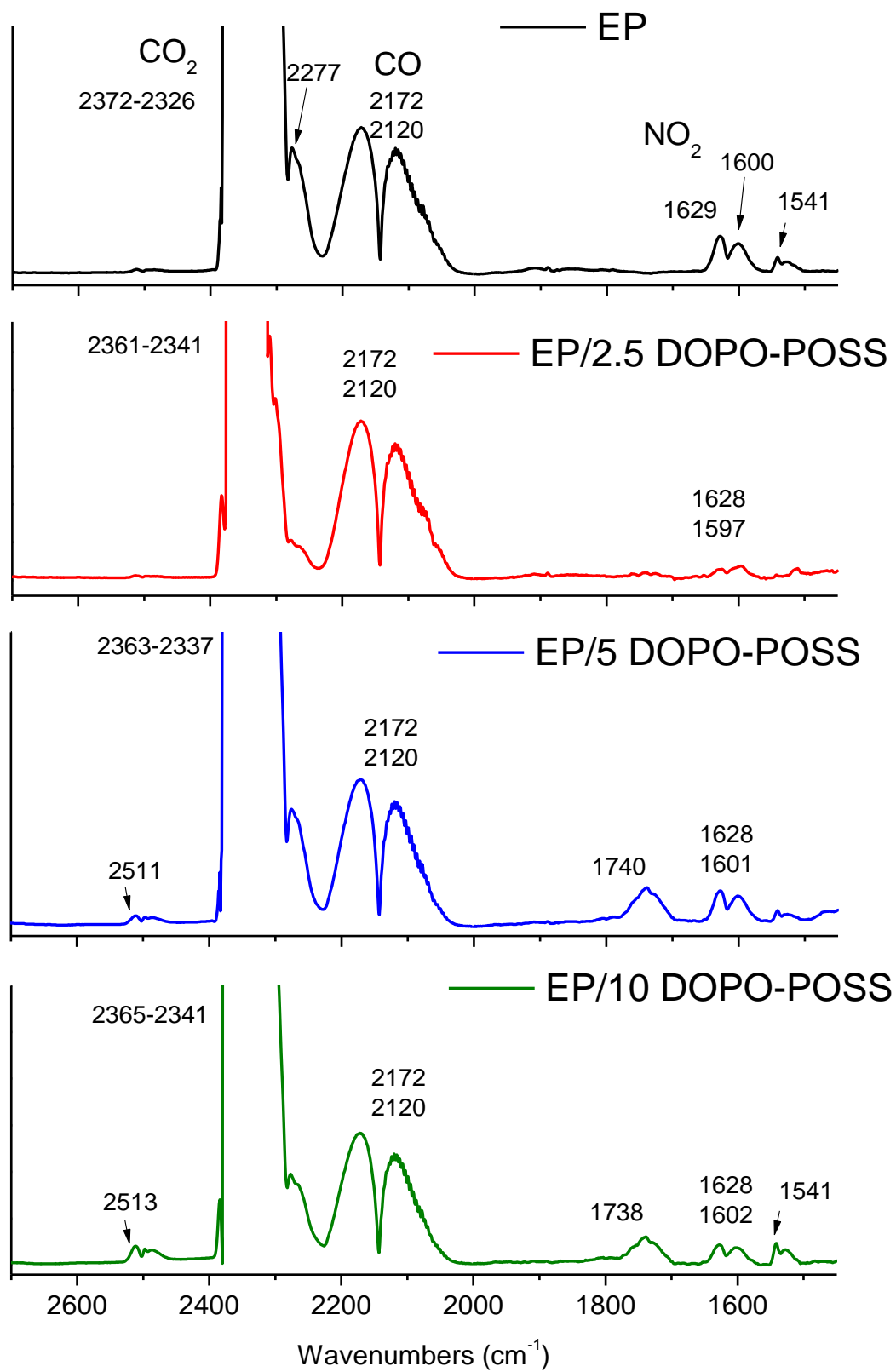

Figure S3: FTIR spectra of gas of EP and EP/DOPO-POSS sampled from the combustion unit: region 2700-1450 cm$^{-1}$

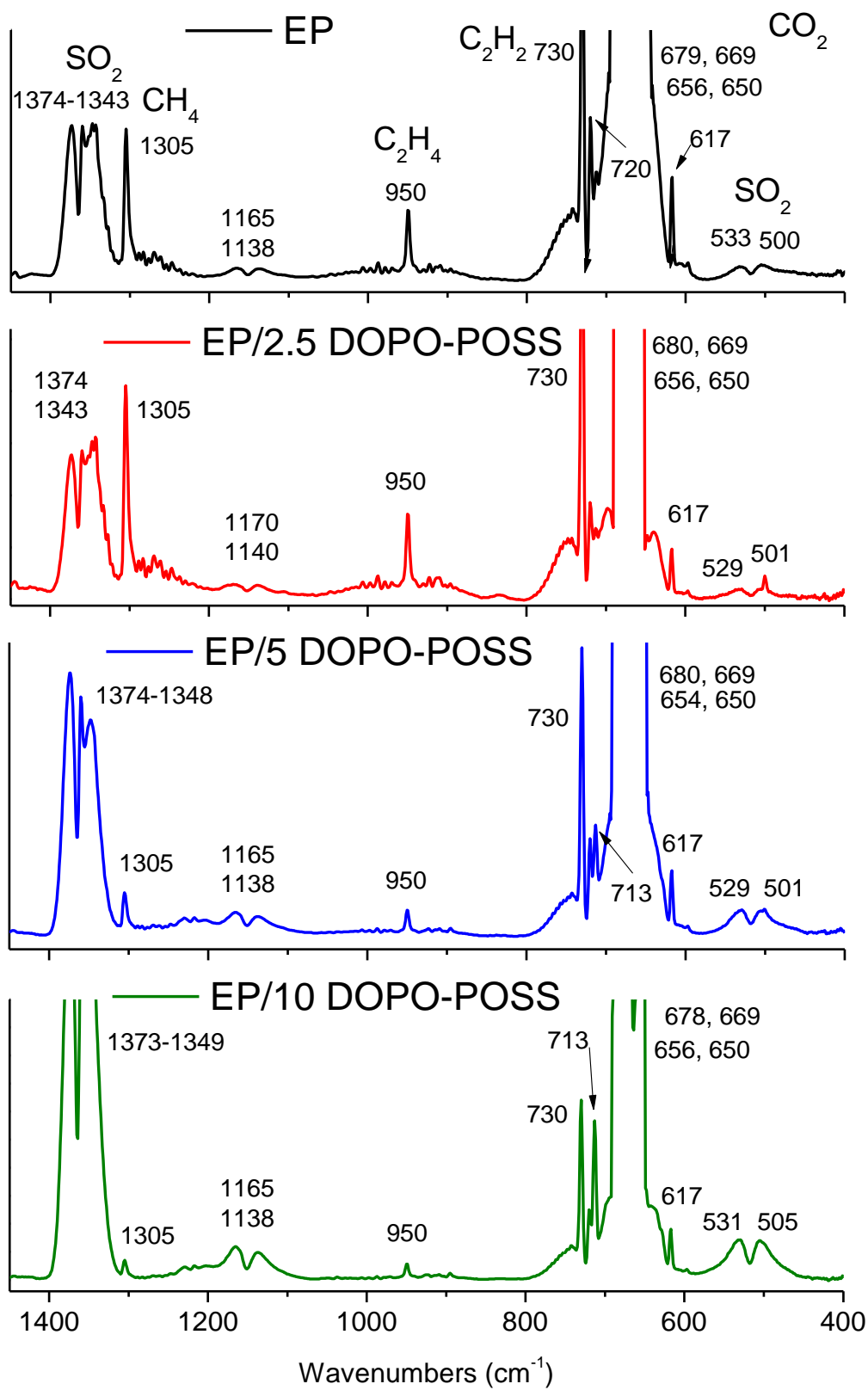

Figure S4: FTIR spectra of gas of EP and EP/DOPO-POSS sampled from the combustion unit: region 1450-400 cm$^{-1}$

Table S1: Flammability parameters for the main products evolved during decomposition of EP and E/DOPO-POSS

| Substance | Boiling Point [°C] | Enthalpy of vaporization [KJ/mol] | Enthalpy of combustion (lower) [KJ/mol] | Lower Flammability Limit [vol%] | Upper Flammability Limit [vol%] | Flash point [°C] | Auto Ignition Temperature in air [°C] |
|---|---|---|---|---|---|---|---|
| **Benzene** | 80 | 34 | 3135 | 1.3 | 7.9 | -11 | 580 |
| **Phenol** | 182 | 59 | 2921 | 1.8 | 8.6 | 79 | 715 |
| **Toluene** | 111 | 37 | 3734 | 1.2 | 7.1 | 4 | 480 |
| **Naphtalene** | 218 | 71 | 4990 | 0.9 | 5.9 | 79 | 526 |
| **Benzonitrile** | 191 | 55 | 3522 | 1.3 | 8.0 | 71 | n.a. |
| **Cresol (meta)** | 202 | 60 | 3520 | 1.1 | 7.6 | 86 | 626 |
| **Cumenol** | 212 | n.a. | n.a. | n.a. | n.a. | 108 | n.a. |
| **Styrene** | 145 | 44 | 4219 | 1.1 | 6.1 | 31 | 490 |
| **Bisphenol A** | 251* | 102 | n.a. | n.a. | n.a. | 227 | n.a. |

Data extracted from V. Babrauskas, Ignition Handbook, Fire Science pubblishers, 2003, except enthalpies of vaporization, form NIST database (http://webbook.nist.gov/chemistry/). Enthalpies are reported in standard conditions. n.a. = not available  *=at 0.017 bar, from NIST database

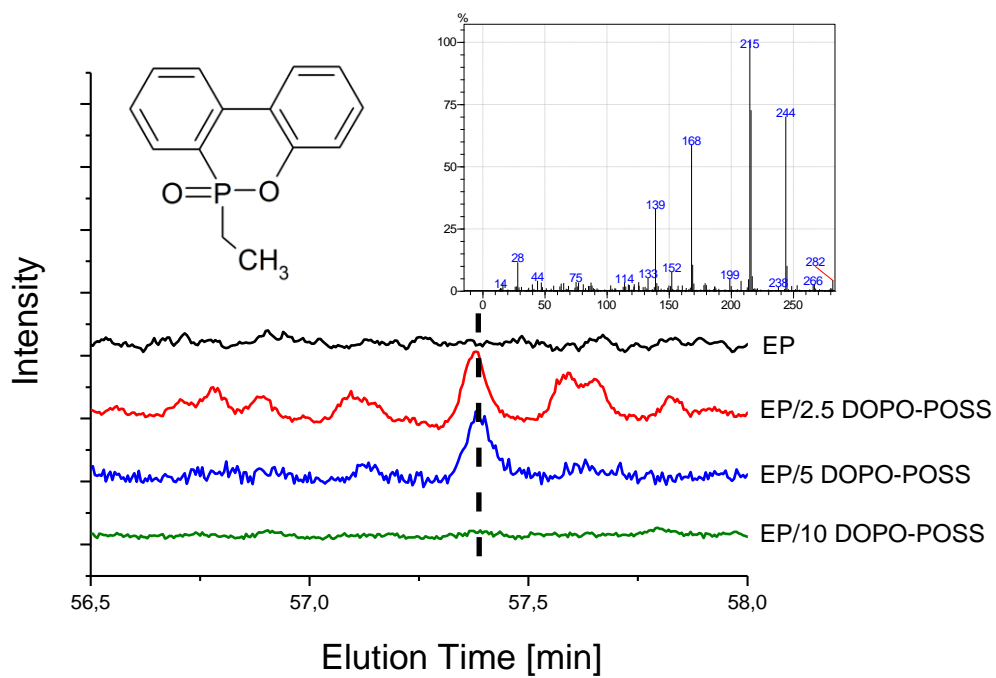

Figure S5: GC evidence for the presence of Ethyl DOPO in the gas phase from combustion of EP/2.5DOPO-POSS and EP/5 DOPO-POSS. Inset shows fragmentation mass spectra. Ethyl DOPO was identified by comparison with DOPO fragmentation mass spectra